\documentclass[twocolumn,a4paper]{nveart}
\usepackage{graphics,pifont,amsmath}

\begin{document}

\begin{frontmatter}
\title{Physics Aspects of the CERN Heavy-Ion Programme \thanksref{qfthep}}
\thanks[qfthep]{Invited talk presented at QFTHEP'2000, Tver, Russia.}

\author{Nick van Eijndhoven}
\address{Department of Subatomic Physics, Utrecht University/NIKHEF\\
         P.O. Box 80.000, NL-3508 TA Utrecht, The Netherlands\\
         email~: nick@phys.uu.nl}

\collab{~}

\begin{abstract}
Statistical calculations within the Standard Model indicate that at extremely
high densities the quarks and gluons will become deconfined, leading to a new state
of matter, the so-called Quark-Gluon Plasma (QGP).
Recently it was announced at CERN that compelling evidence has been obtained
from experimental data that indeed the formation of a deconfined state has been achieved in
very energetic collisions of heavy nuclei.\\
In this presentation I will provide an overview of the main aspects of heavy-ion physics
and will address the various observations which indicate the possible creation of a
deconfined state.
To enable a systematic study of the QGP state, a combined measurement of the various
observables is needed. It will be shown how this can be realised within the ALICE
experiment at the future LHC collider.
\end{abstract}

\end{frontmatter}

\section{Introduction}
The use of ultra-relativistic heavy-ion beams is well suited for studying 
nuclear matter under extreme conditions.
Based on a statistical treatment of QCD \cite{latqcd}, it is expected that
a deconfined state of quarks and gluons, the so-called quark gluon plasma
(QGP), will be created.
The formation, detection and systematic study of such a QGP state would 
yield new information on strong interaction dynamics.

An extensive experimental programme has been undertaken at the CERN-SPS
accelerator complex with sulphur and lead ion beams at beam energies
of 200~GeV and 158~GeV per nucleon, respectively, to search for and investigate QGP formation.
Several observations, such as suppression of the $J/\psi$ resonance \cite{psi}
and enhanced production of strange hadrons \cite{strange},  hint at an interesting
new behaviour of the matter produced in these nucleus-nucleus collisions.

While these observations lead to the perception that the initial phase of the
collision consisted of a hot and dense system with strong rescattering,
consistent with the assumption that a quark gluon plasma was formed, 
a direct signature of the plasma and its properties is still missing.\\
To investigate plasma formation in a more direct way, one needs penetrating
probes which reflect the hot early stage of the interaction. Promptly produced
thermal photons and lepton pairs are generally believed to be able to provide
information about this very early stage.

Combination of the information obtained from penetrating probes with the
various other indirect signals provided via hadronic observables have led to
the conclusion that the data obtained from the CERN-SPS heavy-ion programme
constitute compelling evidence that the onset of deconfinement has been achieved,
as will be outlined below.

\section{Direct photons}  
\begin{figure}[htb]
\begin{center}
\resizebox{4cm}{!}{\includegraphics{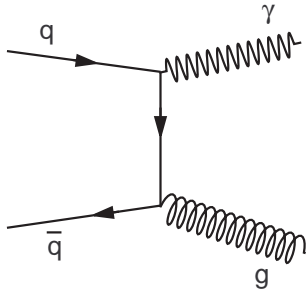}}
\end{center}
\caption{Direct photon production by $q\bar{q}$ annihilation.}
\label{fig:annih}
\end{figure}

\begin{figure}[htb]
\begin{center}
\resizebox{4cm}{!}{\includegraphics{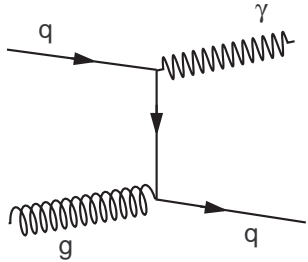}}
\end{center}
\caption{Direct photon production by QCD Compton scattering.}
\label{fig:compton}
\end{figure}

The thermal radiation of a deconfined system is a source of
direct photons, i.e. photons not originating from hadron decays.
Direct photons are thought to provide an excellent means for studying
the state of nuclear matter at the various stages of the interaction, since photons
decouple from the nuclear system immediately after their 
production and are essentially uninfluenced by the hadronisation process.
The various sources from which direct photons may be produced are :

\begin{itemize}
\item Hard QCD processes, i.e. initial interactions of the constituents.
\item Thermal radiation from a Quark Gluon Plasma (QGP),
      mixed Quark Gluon/hadron phase or from a pure hadron gas.
\end{itemize}
The main processes contributing to direct photon production are indicated in
Figs.~\ref{fig:annih} and \ref{fig:compton}. 

In the case of a radiating hadron gas the relevant processes for photon
production, in addition to the familiar decays of $\pi^{0}, \eta, \omega, ...$,
are~:

\begin{itemize}
\item $\pi\pi \rightarrow \rho\gamma \quad$ (Annihilation)
\item $\pi\rho \rightarrow \pi\gamma \quad$ (Compton scattering)
\item $\rho \rightarrow \pi\pi\gamma \quad$ (Decay)
\end{itemize}
In this case the diagrams for annihilation and Compton scattering are
obtained by replacing the (anti)quarks by pions and the gluons by
$\rho$ mesons in the Figs.~\ref{fig:annih} and \ref{fig:compton}.

Theoretical calculations indicate that the above processes dominate in 
different transverse momentum ($p_{\perp}$) regions, of which
the high $p_{\perp}$ (or hard) regime ($p_{\perp} \geq 3$ GeV/c)
is well understood and can be described by perturbative QCD.
The low $p_{\perp}$ (or thermal) regime, however, is not understood
theoretically nor experimentally established.

Photon emission in the thermal regime is dominated by the decay of 
neutral mesons (mainly $\pi^0 \rightarrow \gamma\gamma$ and 
$\eta \rightarrow \gamma\gamma$) produced in the latest stage of the
reaction.
These hadronic decay photons yield a large background which hinders
the search for a possible direct photon signal.

\begin{figure}[htb]
\begin{center}
\rotatebox{90}{\resizebox{6.5cm}{!}{\includegraphics{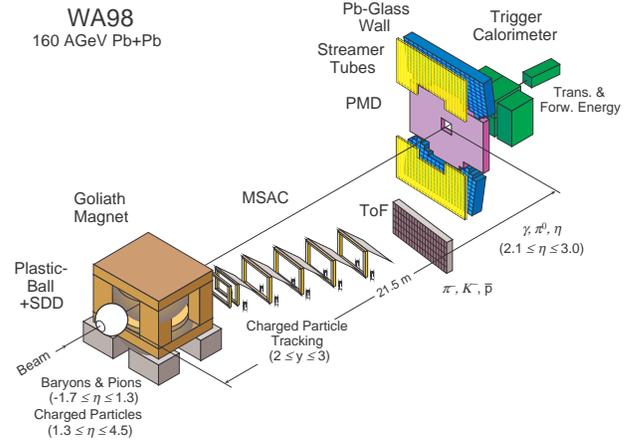}}}
\end{center}
\caption{Experimental setup of the WA98 experiment.}
\label{fig:wa98setup}
\end{figure}

The CERN-SPS experiment WA98 \cite{wa98tp}, shown in Fig.~\ref{fig:wa98setup},
is optimised for photon detection. The main components concerning the study of
direct photon production are the leadglass electromagnetic calorimeter system and
the streamer tube detectors, which serve as a charged particle veto.
The centrality of the collisions is derived from the information obtained from the ZDC
and MIRAC trigger calorimeters, measuring the forward and transverse energy flow, respectively.

Within WA98 the background problem of the hadronic decay photons has been overcome
by reconstruction of the parent $\pi^{0}$ and $\eta$ mesons by means of a two-photon
invariant mass analysis. The thus obtained $\pi^{0}$ and $\eta$ momentum spectra
provide the basis for a prediction of the hadronic decay photon spectrum.
Comparison of this calculated spectrum with the actually observed photon spectrum
might reveal an excess of photons due to direct production.
The final result of the WA98 observed direct photon yield is shown below in Fig.~\ref{fig:wa98result}.
Further details about this analysis can be found in \cite{wa98photons}.

\begin{figure}[htb]
\begin{center}
\resizebox{8cm}{!}{\includegraphics{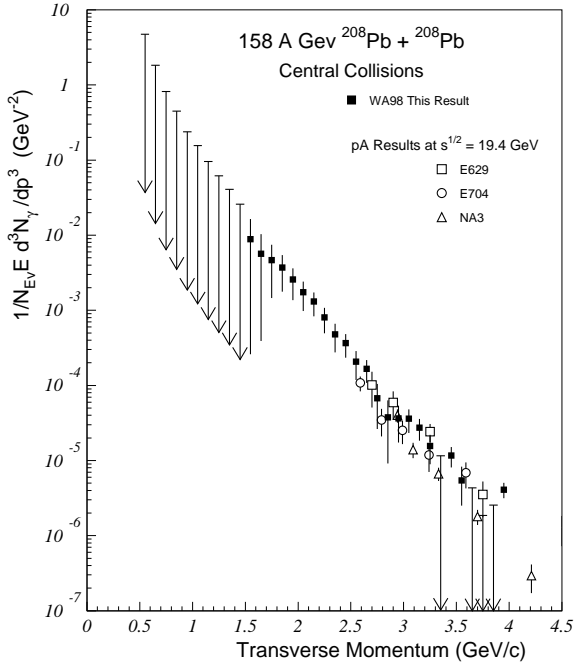}}
\end{center}
\caption{Invariant direct photon yield as observed in central 158~$A$~GeV/c Pb+Pb
         collisions by the WA98 experiment.}
\label{fig:wa98result}
\end{figure}

From the WA98 result it is seen that indeed a direct photon signal is present in the data.
However, due to the limitations imposed by the reconstruction of the parent mesons,
only the relatively high $p_{\perp}$ region can be investigated with sufficient accuracy.
Since in this region direct photons from hard initial scattering processes are expected,
as reflected by the scaled p-A data in Fig.~\ref{fig:wa98result}, no conclusion may be drawn
concerning direct photons originating from thermal radiation.

\newpage

\begin{figure}[htb]
\begin{center}
{\large \bf WA93 preliminary}\\
\resizebox{8.5cm}{!}{\includegraphics{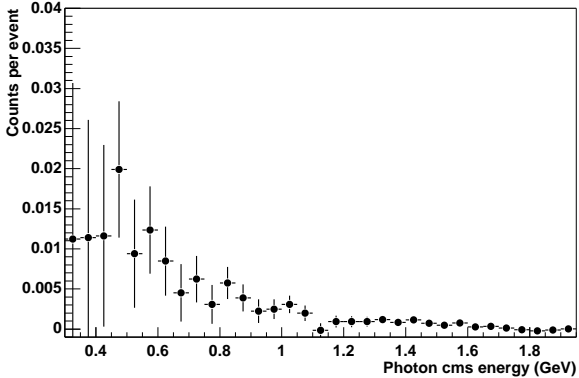}}
\end{center}
\caption{Preliminary result on direct photon production as observed in central 200~$A$~GeV/c S+S
         collisions by the WA93 experiment.}
\label{fig:wa93result}
\end{figure}

In order to investigate the thermal regime, an alternative analysis based on inclusive
photon spectra \cite{nve} is currently in progress.
Application of this method on data obtained earlier with the WA93 experiment \cite{wa93}, an experiment
comparable with WA98 but using 200~$A$~GeV/c S+S interactions at the CERN-SPS instead,
shows a possible direct photon signal due to thermal radiation. 
The preliminary WA93 result is shown in Fig.~\ref{fig:wa93result}.

Theoretical models \cite{tgam} indicate that the temperature of the created system
determines the shape of the direct photon spectrum, whereas the absolute photon yield reflects the
space-time evolution of the radiating system. 
Consequently, observations of direct photon spectra in the thermal regime are expected to
reveal the temperature profile and space-time history of the interaction process.\\
In case of the larger Pb+Pb system, compared to S+S, a relatively enhanced production of thermal
direct photons is expected and thus the outcome of a similar analysis performed on the WA98 data is
eagerly awaited.

\section{Lepton pairs}
\begin{figure}[htb]
\begin{center}
\resizebox{4cm}{!}{\includegraphics{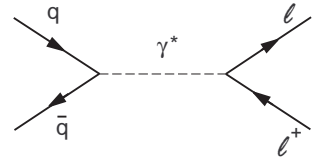}}
\end{center}
\caption{Drell-Yan process for lepton pair production.}
\label{fig:Drell-Yan}
\end{figure}

Another probe to investigate the early stage of the interaction is provided by an invariant
mass analysis of lepton pairs.
One source of di-lepton production is the Drell-Yan process shown in Fig.~\ref{fig:Drell-Yan}. 

\begin{figure}[htb]
\begin{center}
\resizebox{8.5cm}{!}{\includegraphics{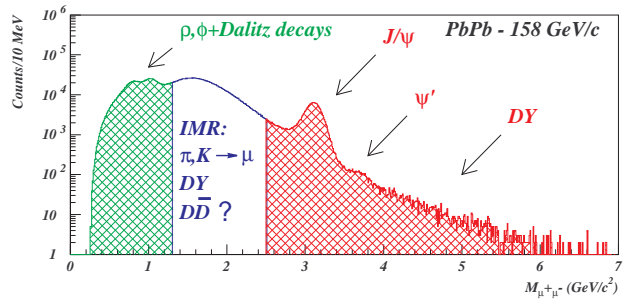}}
\end{center}
\caption{Composition of the di-lepton invariant mass spectrum.
         DY indicates the Drell-Yan domain, whereas the IMR is the
         so-called 'Intermediate Mass Region'.}
\label{fig:imr}
\end{figure}

At sufficiently high energies, the Drell-Yan process is calculable using perturbative
QCD and thus an expression for the invariant mass spectrum of the various lepton pairs
may be obtained. However, due to the di-lepton decay channels of various particles and 
resonances, the lepton pair invariant mass spectrum contains an additional rich structure
at lower energies, as is shown in Fig.~\ref{fig:imr}.

By selecting different domains in the di-lepton invariant mass spectrum, various physics
processes can be addressed. In such a study, the Drell-Yan spectrum may be used
to quantify lepton pair production due to other processes, as explained hereafter.

\subsection{Suppression of the $J/\psi$ resonance}
In the case that in heavy-ion collisions a system of deconfined nuclear matter is created,
it is expected that the members of newly created $q\bar{q}$ pairs within that medium can easily
drift apart and combine with other (anti)quarks from the system to form the final hadrons.
Consequently, a decrease in the yield of resonances like the $J/\psi$ is expected \cite{psith}
compared to interactions in which no such deconfined system is created.

\begin{figure}[htb]
\begin{center}
\rotatebox{90}{\resizebox{!}{8.5cm}{\includegraphics{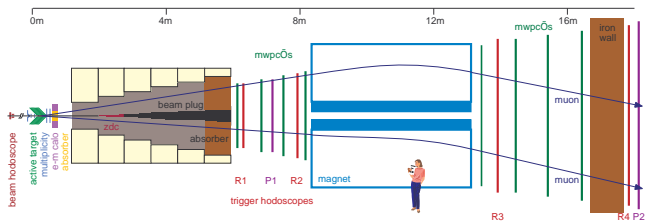}}}
\end{center}
\caption{Experimental setup of the NA50 experiment.}
\label{fig:na50setup}
\end{figure}

The NA50 experiment \cite{na50} at the CERN-SPS is optimised to measure the
production of the $J/\psi$ resonance in heavy-ion collisions by means of the invariant mass
spectrum of muon pairs, originating from the decay channel $J/\psi \rightarrow \mu^{+}\mu^{-}$.
The NA50 experimental setup is shown in Fig.~\ref{fig:na50setup}.

Reconstruction of the tracks of the emerging muons allows the determination of the di-muon invariant
mass spectrum in a domain around the $J/\psi$ peak.  
In addition, measurement of the total forward and transverse energy produced in the interaction
provides information concerning the centrality of the collision.
Consequently, the yield of $J/\psi$ resonances as a function of the centrality can be extracted.

\begin{figure}[htb]
\begin{center}
\resizebox{8.5cm}{!}{\includegraphics{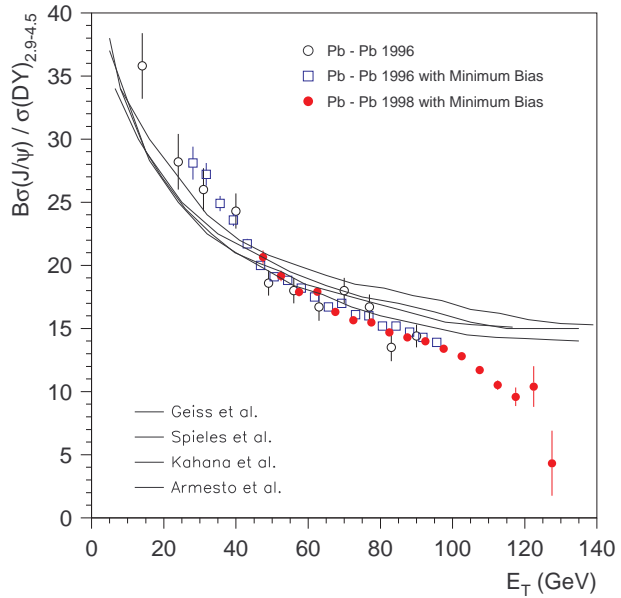}}
\end{center}
\caption{$J/\psi$ resonance production as a function of the total transverse energy
         as measured within the NA50 experiment.
         The curves indicate various theoretical calculations of the expected yields.}
\label{fig:na50result}
\end{figure}

To obtain a normalisation for the $J/\psi$ production, the yield is measured relative to the
Drell-Yan process. However, since the Drell-Yan background cannot be determined very precisely
around the $J/\psi$ invariant mass region, an extrapolation of the Drell-Yan spectrum measured
at higher energies has been used.\\
The final result of this analysis is shown in Fig.~\ref{fig:na50result}, whereas further
details concerning the analysis procedure and references to the theoretical models
can be found in \cite{na50psi}.

From Fig.~\ref{fig:na50result} it is seen that for central (i.e. high $E_{\rm T}$) collisions
an anomalous suppression of the $J/\psi$ is observed compared to theoretical calculations,
whereas for the more peripheral events the agreement between the data and theoretical
curves is rather good.

The obtained result is in line with the expectations for the creation of a quark-gluon plasma,
however, several aspects prevent a definite conclusion on this point.
Firstly, one can argue that the theoretical curves in Fig.~\ref{fig:na50result} are incorrect
for central collisions due to the presence of nuclear effects which are not well understood, 
or that the extrapolation of the Drell-Yan background to this lower invariant mass domain is
not accurate enough. 
In addition, measurement of the yield of protons at mid-rapidity \cite{na49stopping} for
Pb+Pb collisions at the CERN-SPS indicates that in these interactions there is considerable
nuclear stopping. This implies the presence of baryon-rich matter and increased particle
production in the central reaction zone. Consequently, destruction of earlier produced
$J/\psi$ resonances is enhanced and as such the observed suppression may be due to ordinary
scattering of these resonances.

In order to reach conclusive statements concerning plasma formation, it is believed that the
effect has to be studied in relation with other observables and as such a more systematic
investigation is needed. One possibility would be the relative comparison of the production
of $J/\psi$ and $\psi^{\,\prime}$, which would decrease the uncertainty due to the Drell-Yan
extrapolation. Also a combined measurement with for instance direct photon production would
possibly extend our understanding in these processes.

\subsection{Low mass di-electrons}
Its has also been suggested \cite{shifts} that in-medium effects result in modifications
of particle properties such as the mass and width of resonances.
To explore this effect, the NA45/CERES experiment \cite{na45} focuses on the investigation
of electron pairs in the low invariant mass domain around the $\rho^{0}$ peak.
A shift in the $\rho^{0}$ mass would result in a distorted invariant mass spectrum and
might induce a detectable deviation from the known di-electron yields.

\begin{figure}[htb]
\begin{center}
\rotatebox{-90}{\resizebox{!}{8.5cm}{\includegraphics{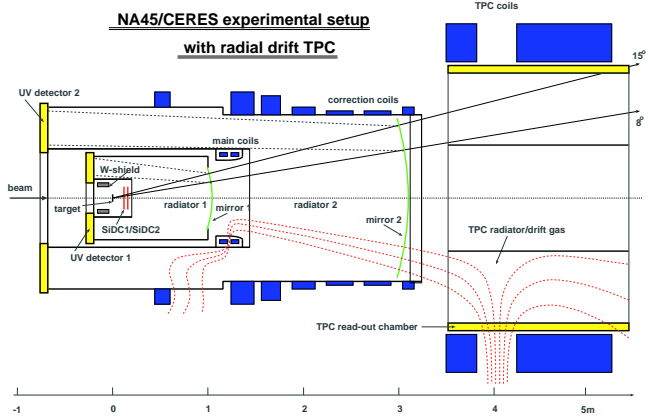}}}
\end{center}
\caption{Experimental setup of the NA45 experiment.}
\label{fig:na45setup}
\end{figure}

The experimental layout of the NA45/CERES experiment is shown in Fig~\ref{fig:na45setup}.
The main components concerning the analysis described here are the silicon trackers,
the Time Projection Chamber (TPC) and the Ring Imaging Cherenkov (RICH) detectors.

The silicon trackers and the TPC provide 3-dimensional space points of the through-going particles
which allows reconstruction of the momentum by means of the observed curvature in the known magnetic
field.
Based on the measured momentum and the observed emission angle of the Cherenkov light in the
RICH detectors, the particle can be identified.  
In this way the momenta of the produced electrons are determined and consequently the
invariant mass spectrum of electron pairs is obtained.

\begin{figure}[htb]
\begin{center}
\resizebox{8.5cm}{!}{\includegraphics{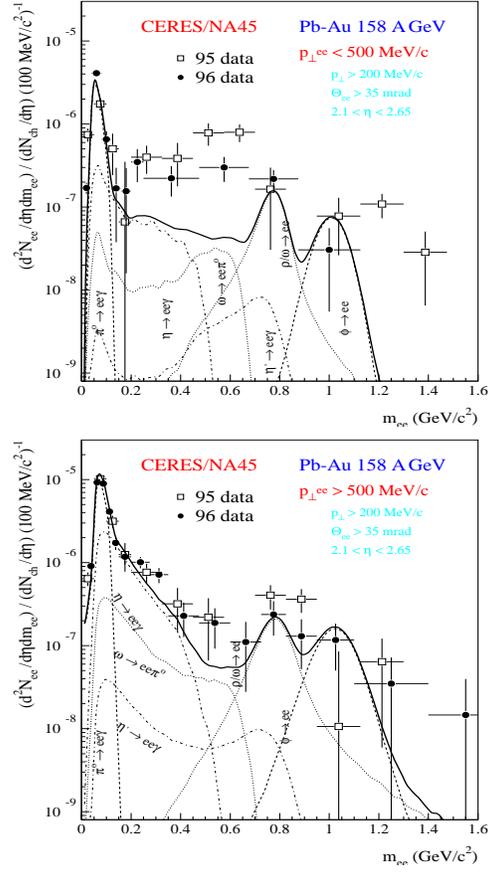}}
\end{center}
\caption{Di-electron invariant mass spectrum for both low (upper panel) and high (lower panel)
         $p_{\perp}$ pairs as measured within the NA45 experiment.
         The curves indicate the various known contributions to the di-electron spectrum.}
\label{fig:na45result}
\end{figure}

The thus obtained invariant mass spectrum is shown in Fig.~\ref{fig:na45result}.
The upper panel shows the data for electron pairs with low transverse momentum ($p_{\perp}<500$~MeV),
whereas the high transverse momentum data are shown in the lower panel.

The high $p_{\perp}$ data are seen to be consistent with the known sources of electron
pair production. However, the low $p_{\perp}$ data indicate an excess of di-electrons
below the $\rho^{0}$ peak, consistent with a shift of the $\rho^{0}$ mass towards a lower value
and broadening of its width.\\
Like before the data are not conclusive concerning deconfinement, since also here an extrapolation
of the Drell-Yan background has been used and again one can question the validity of this extrapolation
over such a large range. 
In addition it would be desirable to have data available for more $p_{\perp}$ selections to
allow a systematic investigation of the $p_{\perp}$ dependence.

In connection to the investigation of in-medium modification of particle properties using
di-lepton probes it is worthwhile to note that these effects can also be explored by
means of hadronic observables.\\
Considering the decay of the $\varphi(1020)$ meson, it is seen that in the decay mode
$\varphi \rightarrow K^{+}K^{-}$ the phase-space is very limited.
A slight mass shift of the $\varphi$ to a lower value might already result in the fact
that this decay channel will become energetically impossible.
Consequently, a combined study of the di-lepton invariant mass spectrum as discussed above
together with an investigation of the kaonic decay mode of the $\varphi$ meson would
increase our understanding of the observed di-electron excess.

\section{Heavy-quark production}
So far the study of the early interaction stage has been discussed on the basis
of a direct investigation using penetrating probes.
However, there exists also an indirect way to assess the early hot phase by means
of hadronic observables.\\
The idea here is that in a deconfined nuclear medium the chiral symmetry may become
(partly) restored \cite{chiral}. This would result in the fact that the mass differences
between the various (anti)quarks decrease. Consequently, a relative enhancement of the
production of heavy quarks is expected in the case of quark-gluon plasma formation.\\
Since this enhanced heavy-quark content has to be reflected in the final state hadrons
leaving the interaction zone, investigation of the 'chemical composition' of the created
system would also enable us to explore the properties of the nuclear matter right after
the initial collision of the nuclei.

\begin{figure}[htb]
\begin{center}
\resizebox{4cm}{!}{\includegraphics{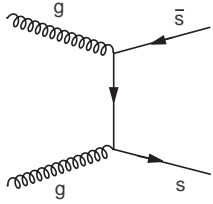}}
\end{center}
\caption{Heavy-quark pair production by gluon fusion.}
\label{fig:gfuse}
\end{figure}

The main process for heavy-quark production is gluon fusion as is indicated in Fig.~\ref{fig:gfuse},
where the case of the creation of an $s\bar{s}$ pair is illustrated.  

Secondary processes like for instance $\pi + n \rightarrow K + \Lambda$ can also lead to
strangeness enhancement when the production process approaches equilibrium conditions.
However, whether or not equilibrium conditions are achieved depends largely on the production
cross sections of the various particles. Thus it is seen that for kaon production indeed
equilibrium conditions might be achieved in nucleus-nucleus collisions, however, this is
very unlikely to happen for the production of (multi)strange hyperons and charmed particles.\\
Therefore, the study concerning a possible heavy-quark production enhancement due to deconfinement
focuses primarily on the yields of (multi)strange hyperons and, if experimentally possible,
charmed particles. 

\begin{figure}[htb]
\begin{center}
\resizebox{6cm}{!}{\includegraphics{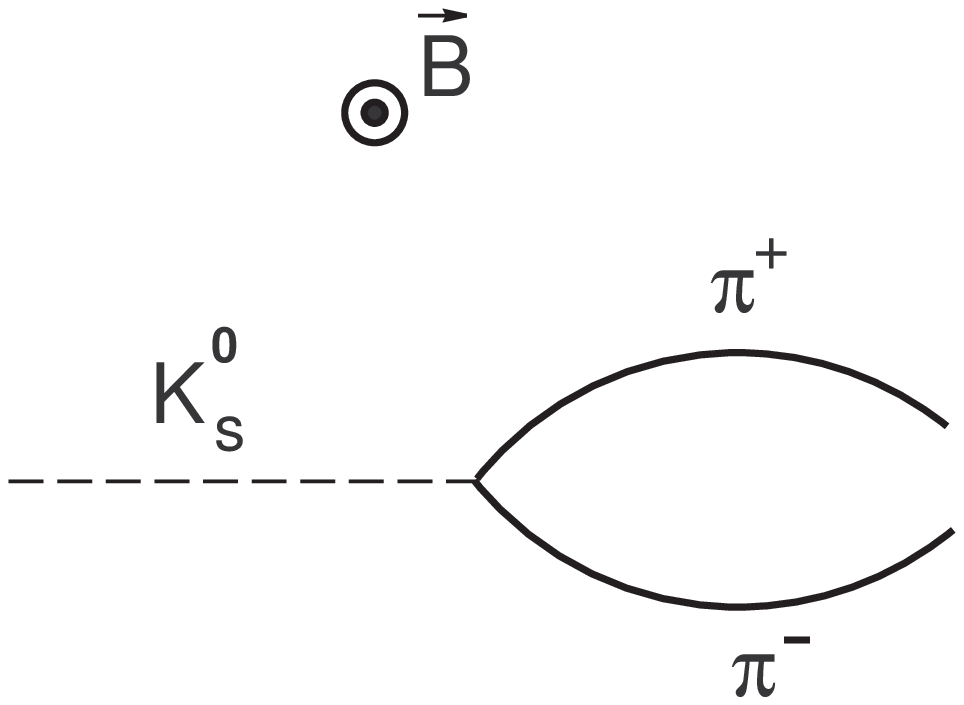}}\\
\resizebox{9cm}{!}{\includegraphics{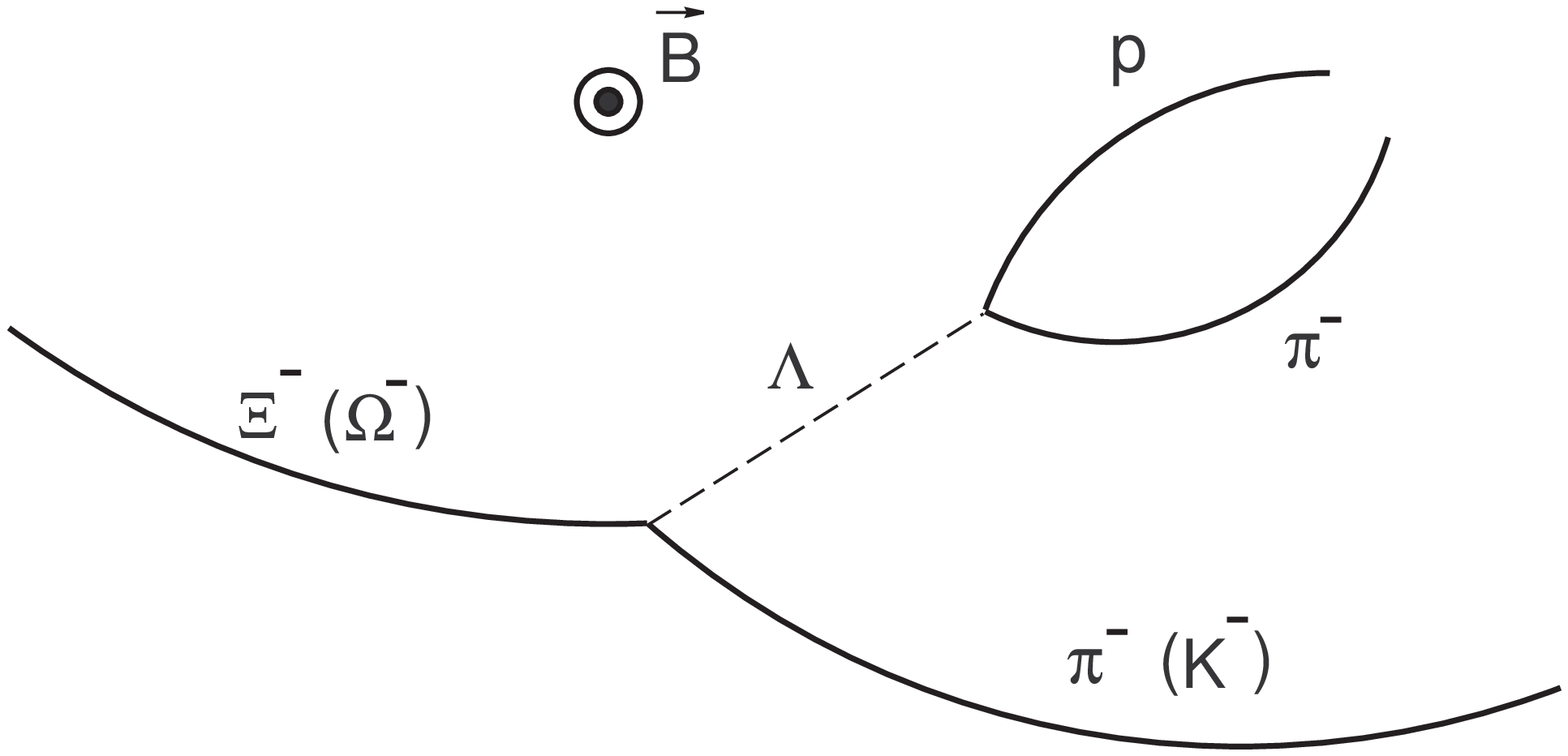}}
\end{center}
\caption{Decay processes of various strange particles.
         The particles indicated between brackets denote the decay process of the
         $\Omega^{-}$ baryon instead of the $\Xi^{-}$ decay.}
\label{fig:kinfit}
\end{figure}

Because of the fact that hadrons containing one or more heavy quarks are rather unstable,
they have to be identified experimentally by means of their decay products.
In Fig.~\ref{fig:kinfit} various decays of strange particles are shown as they would appear
within a magnetic field. From the curvature of the observed charged tracks the particle momenta
can be determined, whereas subsequent kinematical fitting allows identification of the
parent particle with the heavy-quark content. 

\begin{figure}[htb]
\begin{center}
\resizebox{8.5cm}{!}{\includegraphics{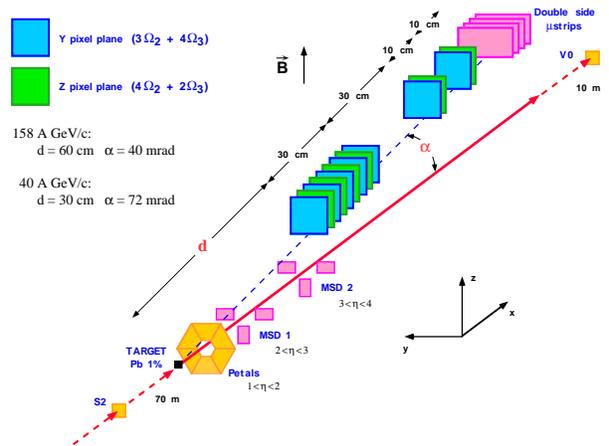}}
\end{center}
\caption{Experimental layout of the NA57 experiment at the CERN-SPS Pb beam facility.}
\label{fig:na57}
\end{figure}

The average decay length of the strange particles amounts to about a few centimeters,
whereas charmed particles have even shorter life times resulting in decay paths in the
order of a millimeter or less.\\
From this it is seen that in order to identify these particles via their decays, the
experimental setup has to be capable of recording the space-points of through-going particles
with a very high precision. One of these experiments is the CERN-SPS NA57 experiment \cite{na57}
of which the experimental layout is shown in Fig.~\ref{fig:na57}. 

The main detector component providing the precise space-point information consists of
a telescope of silicon tracking detectors. Close to the interaction vertex there are
several planes of silicon pixel detectors, whereas further downstream silicon microstrip
detectors provide a leverarm to improve on the measurement of high-momentum tracks.\\
The entire telescope is mounted on a rail which allows coverage of various rapidity ranges
by variation of the angle of inclination. The centrality of the interaction is deduced
from the observed particle multiplicity in the various silicon strip and scintillator panels
upstream of the pixel detectors.
      
\begin{figure}[htb]
\begin{center}
\resizebox{8cm}{!}{\includegraphics{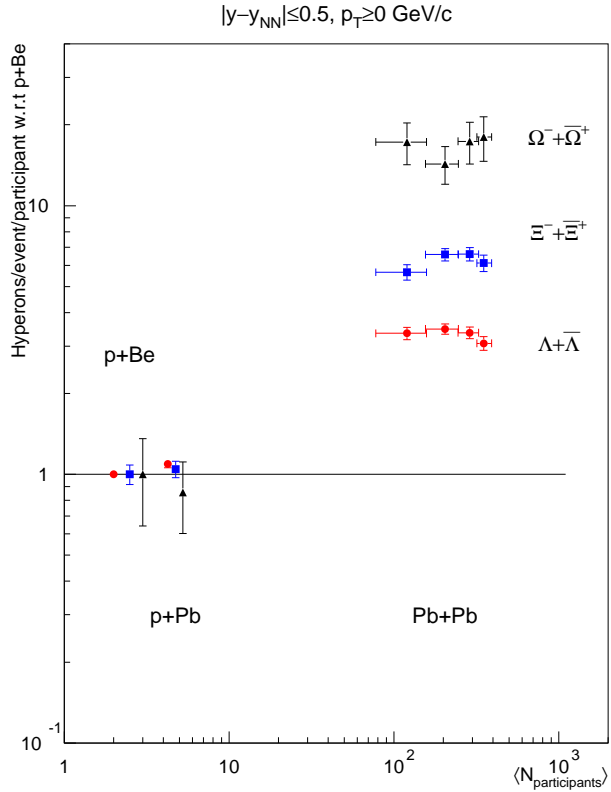}}
\end{center}
\caption{Strange particle yields observed with the CERN-SPS proton and lead beams.
         All yields have been normalised to the p+Be data.}
\label{fig:sresult}
\end{figure}

The above setup was used with the CERN-SPS proton and lead beams and the results \cite{na57result} on
the yields of the various strange particles on different targets are shown in Fig.~\ref{fig:sresult}. 

From the NA57 data it is seen that the higher the strangeness content of the hyperons is,
the stronger the observed enhancement becomes. This observation is in line with the 
expectations in case of deconfinement and cannot be explained in terms of ordinary hadronic
interactions.\\
However, also here the data are not conclusive, since the NA57 measurements assess only a narrow
rapidity region around mid-rapidity and baryon production at mid-rapidity
is strongly influenced by the amount of nuclear stopping in the collision of the nuclei.
In order to make strong statements about (the onset of) deconfinement based on the observed
strangeness enhancement, a systematic study of a possible rapidity dependence of the observations
and the effect of nuclear stopping is essential.

\begin{figure}[htb]
\begin{center}
\resizebox{8cm}{!}{\includegraphics{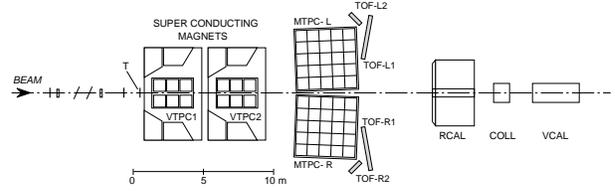}}
\end{center}
\caption{Experimental layout of the NA49 large acceptance multi-purpose experiment.}
\label{fig:na49}
\end{figure}

A first step towards a more systematic investigation over a large rapidity coverage is provided
by the NA49 multi-purpose experiment \cite{na49}, of which a schematical setup is shown in
Fig.~\ref{fig:na49}
 
Also in this setup we recognise a calorimeter system to measure the forward and transverse
energy flow in order to determine the centrality of the collision.
The main component, however, consists of a system of four TPCs which provide the 3-dimensional
space points of the through-going charged particles over a large fraction of the
phase-space.

\begin{figure}[htb]
\begin{center}
\resizebox{8cm}{!}{\includegraphics{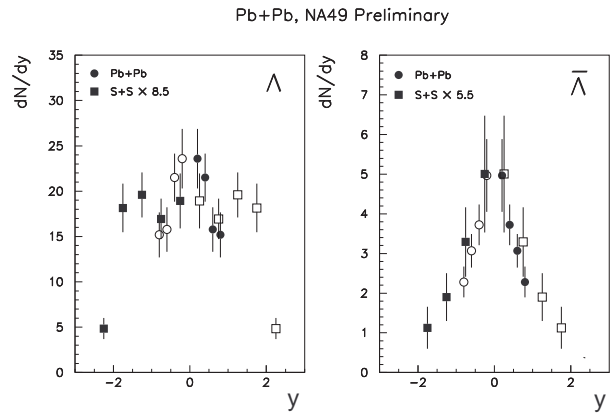}}
\end{center}
\caption{Rapidity distribution of (anti)lambdas as observed within the NA49 experiment.}
\label{fig:na49lambdas}
\end{figure}

Within this experiment the production of (anti)lambda hyperons has been studied \cite{na49lambdas} 
for both sulphur and lead beam induced interactions. The results are shown in Fig.~\ref{fig:na49lambdas}
 
From these data it is seen that the $\bar{\Lambda}$ yield is concentrated around mid-rapidity for
both the sulphur and lead beam data, whereas in the case of $\Lambda$ production the sulphur
induced data show a more flat rapidity distribution than the lead beam data.\\
A plausible explanation for this is the fact that the nuclear stopping in the case of the
lead beam data is much stronger than in the case of the sulphur interactions. This increased
stopping would result in an enhancement of the amount of nucleons at mid-rapidity and thus
lead to an increased $\Lambda$ production in the central region.
Indeed an enhanced yield of protons has been observed \cite{na49stopping} in the case
of lead beam induced interactions, compared to the sulphur data.

\begin{figure}[htb]
\begin{center}
\resizebox{8cm}{!}{\includegraphics{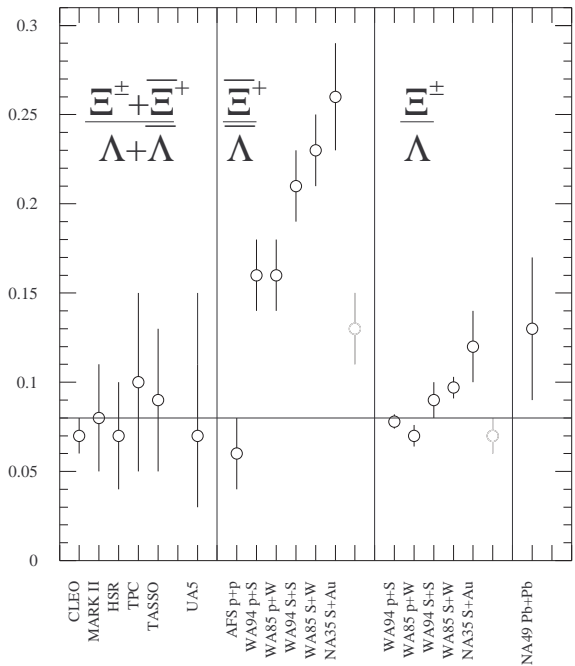}}
\end{center}
\caption{Relative (anti)hyperon yields as observed within various experiments.}
\label{fig:na49ratios}
\end{figure}

This observation might also provide a clue to the fact that in the data of experiments
probing only a small interval around mid-rapidity \cite{na49ratios} the yield of anti-hyperons
was seen to be enhanced compared to the hyperons, as shown in Fig.~\ref{fig:na49ratios}.

From the above it is seen that indeed a systematic investigation of rapidity dependence in
the studies of particle yields is badly needed.
In addition, rapidity dependence of the observed strangeness enhancement could provide
information about the degree of chemical equilibrium reached in the nucleus-nucleus collisions. 
In the case of a fully chemical equilibrated system, no rapidity dependence is expected over
a large region around mid-rapidity.

\begin{figure}[htb]
\begin{center}
\resizebox{8cm}{!}{\includegraphics{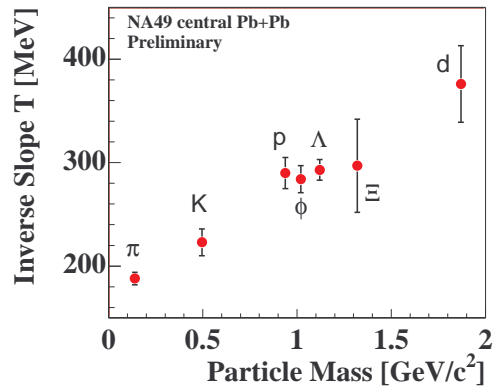}}
\end{center}
\caption{Inverse slopes of the transverse momentum spectra of various particle species
         as observed within the NA49 experiment.}
\label{fig:na49slopes}
\end{figure}

Another feature of the large acceptance multi-purpose NA49 experiment is the capability
of probing the interaction dynamics by means of an investigation of transverse momentum
spectra for a wide range of particle species.\\
The idea here is that in the case of a thermally equilibrated system, the observed spectra
of the emerging particles after hadronic freeze-out should exhibit a thermal character.
This in turn would result in exponential transverse momentum distributions of which the inverse
slope reflects the temperature of the system \cite{ptslope}.\\
Indeed, exponential behaviour of the transverse momentum spectra is observed \cite{na49slopes}
and the inverse slopes as observed within the NA49 experiment are shown in Fig.~\ref{fig:na49slopes}.

From this analysis it is seen that the data are indeed compatible with a thermally equilibrated
system and that the inverse slopes increase with increasing particle mass.\\
One explanation of the observations is that the various particle species freeze-out from
the system at different temperatures. 
Another interpretation of the data can be found in the fact that collective motion of the
produced particles due to a transverse expansion of the created system would also lead to the
observed behaviour \cite{heinz}.
The latter seems to be supported by the observed transverse mass dependence of
the extracted system size by means of an analysis of Bose-Einstein interferometry effects \cite{hbt}.

However, a recent analysis \cite{mypt} of the above results within the phenomenological framework
of particle production via string fragmentation \cite{strings}, has indicated that the observed
behaviour can also be explained without the assumption of an equilibrated system.
In such an interpretation of the data it becomes questionable whether it is allowed to compare
the inverse slopes of mesons and baryons, as was done in the analysis described above, because
of the difference in the production processes of these particles.\\ 
So, once again the data don't provide conclusive results concerning the formation of an
equilibrated system in nucleus-nucleus collisions and a more detailed investigation in
combination with additional observables is needed. 

In the effort to come to conclusive evidence that the observed strangeness
enhancement is due to deconfinement effects, an investigation of a similar enhancement of
charmed particles would provide valuable information.\\
A first attempt to such an analysis has been made within the NA50 experiment, of which
the experimental layout has been described before.

\begin{figure}[htb]
\begin{center}
\resizebox{6cm}{!}{\includegraphics{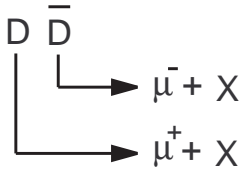}}
\end{center}
\caption{Inclusive muon yield from $D$ meson decay processes.}
\label{fig:ddbar}
\end{figure}

In order to investigate open charm production, the attention is focused to muon pairs
with an invariant mass in the so-called intermediate mass region as indicated
in Fig.~\ref{fig:imr}.
This region is free from resonances and the only sources discussed so far which would
contribute to muon pairs are the Drell-Yan process and random combinations of muons
from pion and kaon decays.
However, the decays of $D$ mesons would result in an additional muon yield, as indicated
in Fig.~\ref{fig:ddbar}.

\begin{figure}[htb]
\begin{center}
\resizebox{8.5cm}{!}{\includegraphics{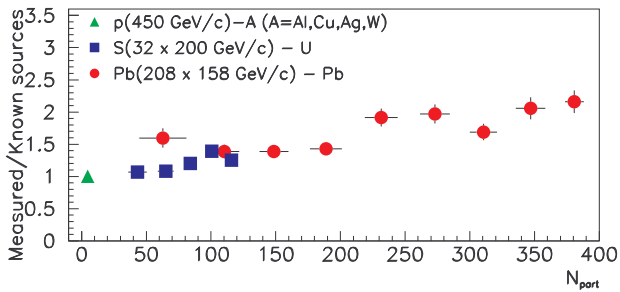}}
\end{center}
\caption{Di-muon yield in the intermediate mass region as observed within the NA50
         experiment. The data have been normalised to the theoretical calculations as
         oulined in the text.}
\label{fig:na50charm}
\end{figure}

These decay processes would obviously give rise to an increase in the yield of muon pairs
and as such a possible di-muon excess above theoretical calculations could indicate an
enhancement of charmed quark production.\\
Using the PYTHIA \cite{pythia} Monte Carlo program and a proper scaling to describe nucleus-nucleus
collisions, a complete description of the di-muon yield in the intermediate mass region due to
known processes is obtained. In Fig.~\ref{fig:na50charm} the di-muon yield in the intermediate
mass region is shown as a funcion of the number of participating nucleons as observed within the
NA50 experiment \cite{na50charm}. The data have been normalised to the theoretical calculations.    

It is seen that the proton induced data and the peripheral heavy-ion data are consistent
with the theoretical calculations, whereas the central heavy-ion data clearly show an excess.
The observed excess is in line with the expectations in case of the creation of a deconfined
system. Furthermore, the transverse momentum spectrum of the excess di-muon pairs is found
to be incompatible with muon pairs from the Drell-Yan process or random combinations of
muons from pion and kaon decays \cite{na50charm}.\\
However, a more conclusive result concerning enhanced $D$ meson production would need the
possibility to reconstruct the different production vertices of the muons indicated in
Fig.~\ref{fig:ddbar}. Unfortunately this is not possible with the current NA50 experimental
setup and consequently an experiment with enhanced secondary vertex reconstruction potential
is urgently needed. 

\section{The future ALICE experiment}
From the above it is seen that the data from the CERN-SPS heavy-ion programme exhibit
characteristics that are compatible with (the onset of) deconfined nuclear matter.
However, the various signatures have been observed separately in different experiments
and to come to conclusive results, a combined measurement of the various observables
is needed.

\begin{figure}[htb]
\begin{center}
\resizebox{8.5cm}{!}{\includegraphics{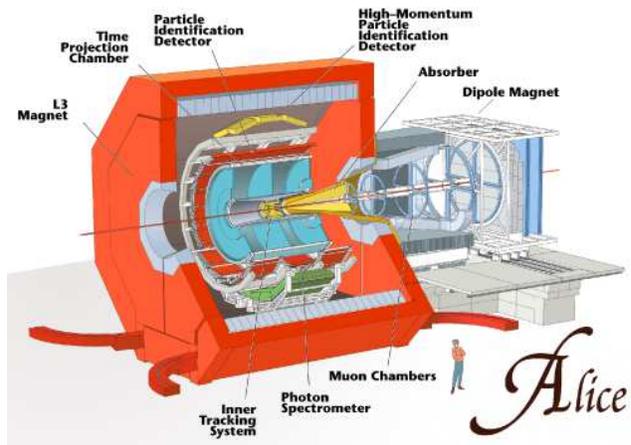}}
\end{center}
\caption{Schematic view of the ALICE detector.}
\label{fig:alicesetup}
\end{figure}

Such a combined measurement will become possible within the ALICE \cite{alicetp} experiment at
the future CERN LHC accelerator, as is outlined hereafter.
A schematic representation of the ALICE detector is shown in Fig.~\ref{fig:alicesetup}.

\begin{figure}[htb]
\begin{center}
\resizebox{8.5cm}{!}{\includegraphics*{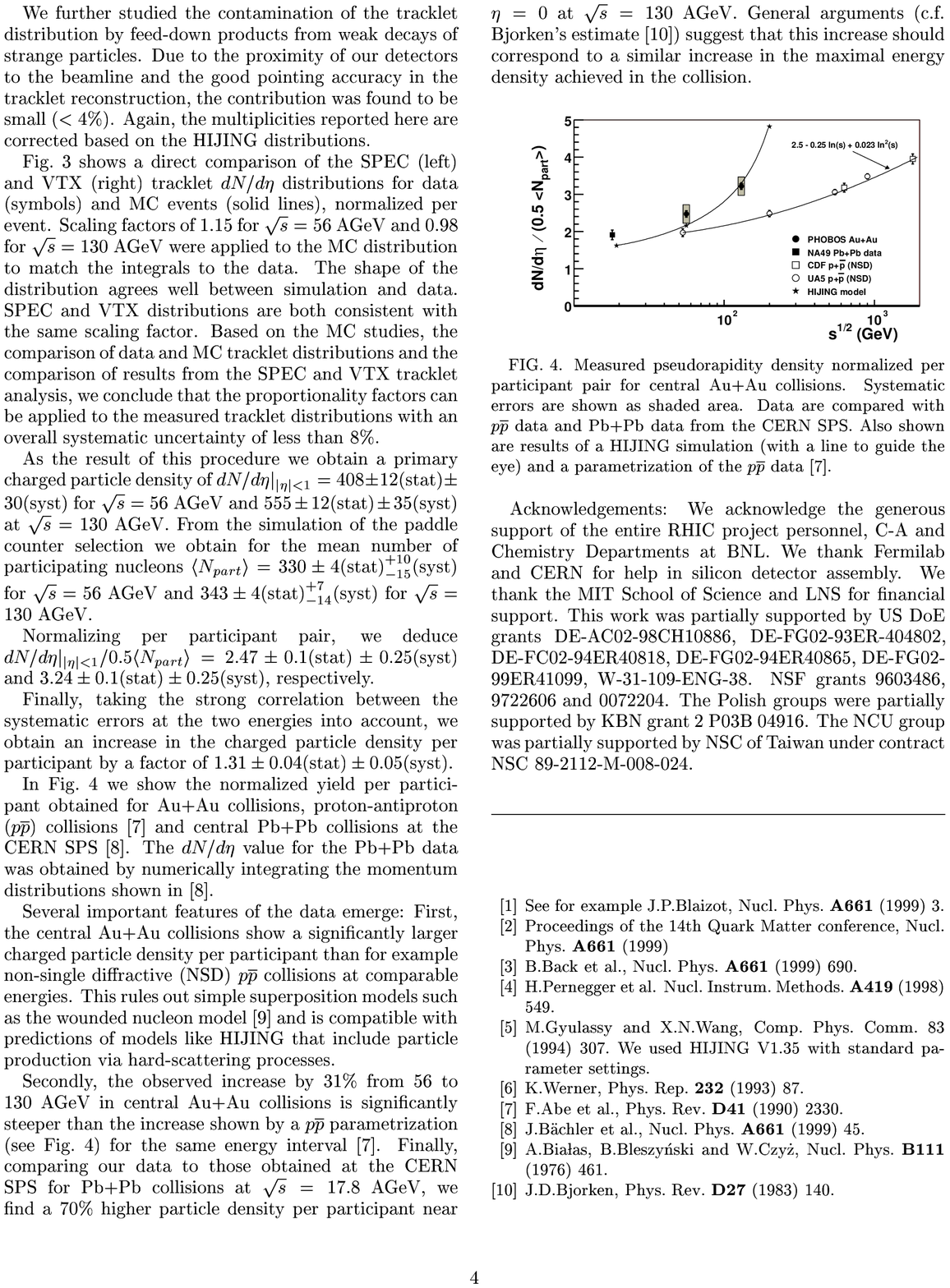}}
\end{center}
\caption{Charged particle multiplicities as measured within PHOBOS at RHIC.}
\label{fig:rhicmult}
\end{figure}

At the LHC, interactions between lead ions will take place at a CMS energy of about 5.5~TeV
per nucleon-nucleon collision. Simulation studies \cite{evtgen} based on various theoretical
models \cite{models} indicate that around mid-rapidity the primary particle yield will
be of the order of 6000 charged particles per unit of rapidity.\\
Recent results of the PHOBOS experiment at RHIC \cite{phobos} have strengthened our believe
in the theoretical predictions. Charged particle multiplicity measurements in the rapidity range
$|\eta|<1$ for collisions between gold ions at CMS energies of 56 and 130~GeV per nucleon-nucleon
collision indicate that the HIJING model agrees rather well with the data, as shown in
Fig.~\ref{fig:rhicmult}.\\
Furthermore, the data in Fig.~\ref{fig:rhicmult} suggest that nucleus-nucleus collisions
at RHIC energies cannot be regarded as a simple superposition of nucleon-nucleon interactions.
The observed increase of the multiplicity in the case of gold-gold collisions compared to the
proton-proton data might indicate that higher energy densities are reached in the created
system of nucleus-nucleus interactions.\\
However, in general particle multiplicities are depending on both the CMS energy of the
interaction and the rapidity of the produced particles. To my opinion an accurate investigation
of the latter is needed before conclusive results can be obtained on basis of analyses
like the one reflected in Fig.~\ref{fig:rhicmult}.

In the ALICE detector design and physics performance studies \cite{alicetp} an upper limit
of 8000 charged particles per unit of rapidity in the central region has been used.\\
Having adopted this safety margin, we feel confident that the detector will be able
to cope with the particle densities produced in the most violent heavy-ion collisions at the LHC.

\subsection{Photon detection}
The photon spectrometer (PHOS) \cite{phostdr}, consisting of 18000 lead-tungstate crystals
situated at a distance of about 5~m from the interaction point, will record the energy of
the emerging photons.
The dimensions of the individual crystals are 2.2~x~2.2~x~18~cm$^{3}$ and
the acceptance of the detector in pseudo-rapidity and azimuthal angle amounts to
$|\eta|<0.12$ and $\Delta\varphi=100^{0}$, respectively.\\
Detector simulation studies have shown that with this design the detector occupancy
will be of the order of 15\%.

In view of a photon and neutral meson analysis as pointed out before for the WA98 experiment,
the spatial ($\sigma_{x,y}$) and energy ($\sigma_{E}$) resolution are crucial parameters.
This can readily be seen by the fact that in the case of neutral meson reconstruction by means
of a two-photon invariant mass analysis, the invariant mass resolution ($\sigma_{M}$)
is given by~:

\begin{equation}
\frac{\sigma_{M}}{M}=0.5\cdot
        \sqrt{
         \frac{\sigma_{E_{1}}^{2}}{E_{1}^{2}} +
         \frac{\sigma_{E_{2}}^{2}}{E_{2}^{2}} +
         \frac{\sigma_{\alpha}^{2}}{\tan^{2}(\alpha/2)}
        } ~~~,
\end{equation} 
where $\alpha$ represents the opening angle between the two photons.

Testbeam results of prototype modules have shown that the values for the spatial
and energy resolution are the following~:

\begin{gather}
\sigma_{x,y}=\left(0.44+\frac{3.26}{\sqrt{E}}\right){\rm mm} \\
\frac{\sigma_{E}}{E}=10^{-2}\cdot\sqrt{1+\frac{9}{E}+\frac{9}{E^{2}}} ~~~,
\end{gather}
from which it is seen that the energy resolution is the limiting factor.\\
The corresponding values in the case of WA98 were

\begin{gather}
\sigma_{x,y}=\left(0.4+\frac{7.1}{\sqrt{E}}\right){\rm mm} \\
\frac{\sigma_{E}}{E}=10^{-2} \cdot \left(1.4+\frac{5.8}{\sqrt{E}}\right) ~~~,
\end{gather}
which indicates that the ALICE photon spectrometer will provide more accurate
results for both the inclusive photon and neutral meson measurements.\\
However, the larger multiplicity in the case of ALICE events compared to WA98
will result in a larger combinatorial background in the invariant mass analysis
of central collisions at the LHC.
The statistical accuracy ($\sigma_{N}$) of the neutral meson yields is given by

\begin{equation}
\frac{\sigma_{N}}{N}=\frac{1}{\sqrt{N}}\cdot
\sqrt{\left[2\left(\frac{S}{S+B}\right)^{-1} - 1 \right]} ~~~,
\end{equation}    
where $S$ and $B$ represent the signal and background, respectively.

In order to reduce the background and increase the purity of the photon sample,
a charged particle veto (CPV) detector will be installed in front of the PHOS.
In addition, fake photon signals due to showering neutral hadrons are largely removed
by means of a shower shape analysis \cite{wa98photons}.
Currently, a design with a conversion layer sandwiched in between two CPV layers is
being investigated. This would allow a further reduction of the neutral
hadron contamination, especially (anti)neutrons, at the expense of a slightly worse
energy resolution.\\
Simulation studies have indicated that for photon energies above 250~MeV a photon
reconstruction efficiency of about 90\% can be achieved together with a purity of the
photon sample which exceeds 99\%.

Based on the above parameters it is seen that the PHOS will provide a momentum
resolution for reconstructed $\pi^{0}$'s and $\eta$'s of 1\% and 10\%, respectively,
in the momentum range of 1-10~GeV/c. In addition the high granularity of the detector
will allow $\pi^{0}$ reconstruction up to momenta as high as 30~GeV/c.\\
The statistical accuracy on the neutral meson yields as a function of transverse momentum
($p_{\perp}$) will amount to 1\% for $\pi^{0}$'s with $p_{\perp}>0.7$~GeV/c and 10\% for
$\eta$'s with $p_{\perp}>2$~GeV/c in the case of 200~MeV/c binning and a total sample
of $10^{7}$ events.  

From the above it is seen that the ALICE photon spectrometer will allow an accurate
measurement of (transverse) momentum spectra of neutral mesons via a two-photon invariant
mass analysis as well as a detailed investigation of the direct photon yield over a
large energy range, including the thermal regime.

\subsection{Strange and charmed particle yields}
As was indicated before, the investigation of heavy-quark production requires
accurate charged particle tracking and vertex reconstruction capabilities.\\
To achieve the required accuracy on the recorded space points, the central part
of the ALICE detector comprises a large cylindrical time projection chamber (TPC) \cite{tpctdr}
and a silicon based inner tracking system (ITS) \cite{itstdr}.
Both tracking systems are located in the uniform 0.2~T magnetic field of the ALICE solenoid
which surrounds all the central detectors. 

The TPC, with an inner radius of 90~cm and an outer radius of 250~cm, covers a range
of $|\eta|<0.9$ in pseudo-rapidity.
In this central area it will provide a space-point resolution of 0.3-2~mm in $r\varphi$
and 0.6-2~mm in $z$, where the highest resolution is achieved at small radial distances.
The two-cluster resolution amounts to about 1~cm, in both $r\varphi$ and $z$.\\
In addition, the TPC will provide a ${\rm d}E/{\rm d}x$ measurement with an accuracy of
better than 10\%, allowing for particle identification up to momenta of several GeV/c.

Simulation studies have shown that the momentum resolution for minimum-ionising particles
amounts to 1.2~\%, whereas for 5~GeV/c electrons this was found to be 5\%.
Prototype algorithms for pattern recognition provide already now a track finding efficiency
exceeding 90\% and recent progress indicates that improvement on this is to be expected.

The ITS, consisting of 6 concentric cylindrical silicon layers, is situated between the
interaction point and the TPC and covers the same pseudo-rapidity interval as the TPC.
The system contains double layers of silicon pixel, silicon drift and silicon
microstrip detectors.\\
The inner cylinder, i.e. the first pixel layer, is at a distance of only 4~cm from the
interaction point, whereas the outer radius, being the outmost silicon microstrip layer,
amounts to 45~cm.\\
The space-point resolution ranges from 12-40~$\mu$m in $r\varphi$ and 70-800~$\mu$m in $z$,
where also here the highest resolution is achieved for the innermost layers.
The system provides a two-track resolution of 0.1-0.3~mm and 0.6-2.4~mm in $r\varphi$ and $z$,
respectively.

Simulations show that, using the ITS as a standalone system, a momentum resolution of
better than 2\% is achieved for particle momenta in the range of 0.1-3~GeV/c.\\
However, using the ITS and TPC as a combined tracking system greatly enhances the reconstruction
capabilities.
The track matching between the TPC and ITS is seen to be about 90\% and the spatial resolution
on the primary vertex position is found to be

\begin{equation}
\sigma_{r\varphi}=15\mu{\rm m} \qquad
\sigma_{z}=\left(7+\frac{260}{\sqrt{\d N_{ch}/\d \eta}}\right) \mu{\rm m} ~.
\end{equation}

The above parameters allow determination of the track impact parameters w.r.t. the
primary vertex with an accuracy better than 100~$\mu$m. 
From this it is seen that reconstruction of the decays of (multi)strange baryons will
not be much of a challenge and that even the investigation of open charm production
via the decay channels $D^{0} \rightarrow K \pi$ and $D^{\pm} \rightarrow K \pi \pi$
will be feasible.

\subsection{Lepton pairs}
The ALICE central tracking system, which has been outlined above, will be surrounded by
a transition radiation detector (TRD) \cite{trdtp} with the same pseudo-rapidity
coverage as the TPC and ITS detection systems.\\
The basic operational principle of the TRD is based on the fact that charged particles
emit X-ray radiation when crossing the boundary between two media with different
refraction indices.
By implementing a design with many layers containing two different media and by making
use of the reconstructed tracks of the tracking system, this detector allows for a very
efficient $e^{\pm}/\pi^{\pm}$ separation for momenta above 500~MeV/c.

The electron detection efficiency amounts to about 90\% for momenta above 1.5~GeV/c
and the pion rejection factor in that case is better than 500.
This implies that this detector will enable us to perform di-electron studies probing
the resonances and continuum regimes as indicated in Fig.~\ref{fig:imr}.\\
Furthermore, the TRD will allow the study of semi-leptonic decays of $D$ and $B$ mesons
by analysis of tracks, reconstructed by the central tracking system, that are not emerging
from the interaction point. Since these particles have short lifetimes, the impact
parameters to the primary vertex of their decay products are generally rather small.
However, the central tracking system provides an accuracy of better than 100~$\mu$m on
these impact parameters, which implies that with the TRD also open charm and bottom
production studies become feasible.

In the forward region, the ALICE experiment contains on one side a muon detector,
the so-called forward muon arm \cite{muontp}, covering the pseudo-rapidity
interval $2.4<\eta<4$. The purpose of this detector is to study muon pairs produced
in the violent nucleus-nucleus collisions in order to investigate the yield of
$J/\psi$ and $\Upsilon$ resonances.\\ 
Because of the location in the forward region, the muons within the acceptance are
sufficiently energetic ($>4$~GeV) such that absorber techniques can be used.
The muons are deflected by a 3~Tm dipole magnet and the space points of the tracks
are recorded by several tracking stations, allowing reconstruction of the muon momenta
followed by an invariant mass analysis of the muon pairs.\\
The triggering is performed by the ALICE forward multiplicity detectors (FMD) situated
at $\eta>4$, whereas the primary vertex position will be provided by the inner tracking system.
Simulation studies have shown that an invariant mass resolution of better than 100~MeV
can be obtained for the muon pairs. This means that the detector will be able to resolve
the yields of both the $J/\psi$ and $\Upsilon$ families and as such investigate in detail
the resonances regime shown in Fig.~\ref{fig:imr}.  

\section{Summary and outlook}
Several experiments at the CERN-SPS heavy-ion programme have separately observed effects
that are consistent with (the onset of) deconfined nuclear matter.\\
Most of the observations can be explained on an individual basis by conventional physical effects
in taking both the experimental and theoretical uncertainties to their limits.
However, explaining all observations together in this way seems to be very hard, if at all
possible. This has led to the official CERN statement in february 2000 that compelling evidence
has been obtained that in nucleus-nucleus collisions at SPS energies deconfinement of nuclear
matter might have been observed.

However, it is generally believed that conclusive results can only be obtained by combined measurement
of the relevant observables related to the creation of a deconfined system.\\
The ALICE experiment at the future CERN-LHC heavy-ion programme will be able to provide
such combined investigations for the violent interactions between lead ions at extremely
high collision energies.\\ 
First results of collisions between gold ions at RHIC indicate that indeed
nucleus-nucleus collisions exhibit additional physics signatures compared to proton-proton interactions.
Based on these observations it is expected that the LHC energy domain will enable the exploration
of a completely new physics regime.   

\begin{ack}
The author would like to thank the organisers of QFTHEP'2000 for the warm welcome,
the hospitality and creating a stimulating atmosphere.\\
He also wants to express his gratitude to his colleagues of the various CERN heavy-ion
experiments for providing the information which made this presentation possible.
\end{ack}

\end{document}